# CONDENSATION AND SUBLIMATION OF THIN AMORPHOUS ARSENIC FILMS STUDIED BY ELLIPSOMETRY

## Chikichev S.I.[1,2], Vasev A.V.[1]

[1]Institute of Semiconductor Physics (Novosibirsk, Russia)
[2]Novosibirsk State University (Novosibirsk, Russia)
*vasev@isp.nsc.ru*

**ABSTRACT.** From *in-situ* ellipsometric measurements during condensation of $As_4$ molecular beam on GaAs(001) substrate the arsenic sticking coefficient was determined. For the substrate with Ga-rich surface reconstruction at 60°C the sticking coefficient was found to be $7\times10^{-3}$, whereas for the surface of amorphous arsenic this coefficient was only $8\times10^{-4}$. Kinetic studies of As sublimation were also performed and activation energy for the process was obtained. For amorphous As films grown from $As_2$ beam the activation energy was found to be $\Delta E=1.84\pm0.03$ eV, which is 0.39 eV higher than the enthalpy of *solid-vapor* phase transition. Temperature dependence of evaporation coefficient for amorphous As was determined in the 230÷290°C range.

## INTRODUCTION

Thin layers of amorphous arsenic (a-As) are widely used in GaAs technology and surface studies as protective layers for as-grown surface during ambient transfer of samples from molecular beam epitaxy (MBE) chamber to other vacuum-based experimental set-ups.

This films, usually known as As-cap layers, provide a simple and effective method for preparation of atomically clean surfaces, which are obtained by sublimation of the protective layer (with all adsorbed contaminations) during heating of the capped sample in vacuum to relatively low temperatures (~300°C). Despite the fact that this technique is used by many for a long time [1-4] surprisingly little is known about the kinetics of amorphous As condensations and sublimation on GaAs substrates. This is because the reflection high-energy electron diffraction (RHEED), which is an excellent technique for monitoring epitaxial (i.e. single-crystal) growth, becomes useless for amorphous layers. In the present work we have used *in-situ* ellipsometry to obtain kinetic data during growth and sublimation of thin a-As layers on GaAs substrates.

## EXPERIMENTAL

All experiments were performed in ultra-high vacuum (UHV)

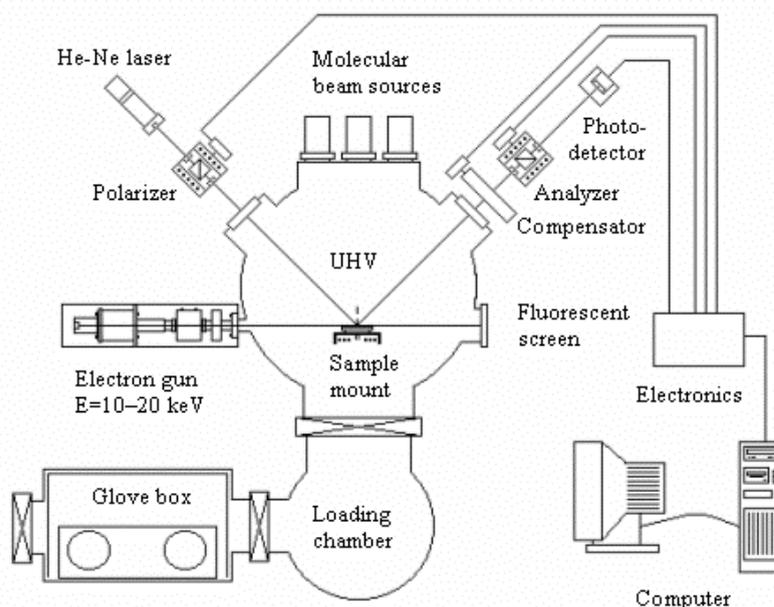

**Figure 1** Schematical representation of experimental set-up.

chamber with base pressure ~$1\times10^{-9}$ torr equipped with automatic single wavelength ellipsometer LEF-701, RHEED gun and $As_4$ molecular beam source, as shown in Fig.1.

The He-Ne laser was used as a light source ($\lambda$=6328Å) and ellipsometric parameters $\psi$ and $\Delta$ were measured with an accuracy of 0.01° and 0.1°, respectively. The incidence angle of light beam on the sample surface was 71.95±0.05°. Accelerating voltage of electron gun was 17kV. Sample temperature was measured by chromel-alumel thermocouple imbedded into molybdenum substrate carrier. Wafers were fixed on the carrier by molten indium providing good thermal contact between sample and Mo-block.

Atomic force microscopy (AFM) was used to study the surface micromorphology on decapped samples. AFM measurements were performed in air at 20°C.

**CONDENSATION**

The substrates used were GaAs wafers cut from bulk undoped single crystals grown by Czochralski technique. They were chemo-mechanically polished on both sides and (001)-oriented within 0.5°.

For substrate preparation we have used a procedure described in [5]. After degreasing in boiling toluene the samples were subjected to polishing etch in $H_2SO_4:H_2O_2:H_2O$=6:1:1 at 60°C and then rinsed in running deionized water. The final chemical treatment was done in a glove box made of organic glass and filled with dry nitrogen. This treatment consisted in room temperature etching of the surface in saturated solution of HCl in ultra-high purity isopropyl alcohol during 3 minutes followed by rinsing in isopropanol. After fixing the wafer on Mo sample holder by means of molten In the substrates were transferred into loading chamber without contact with air.

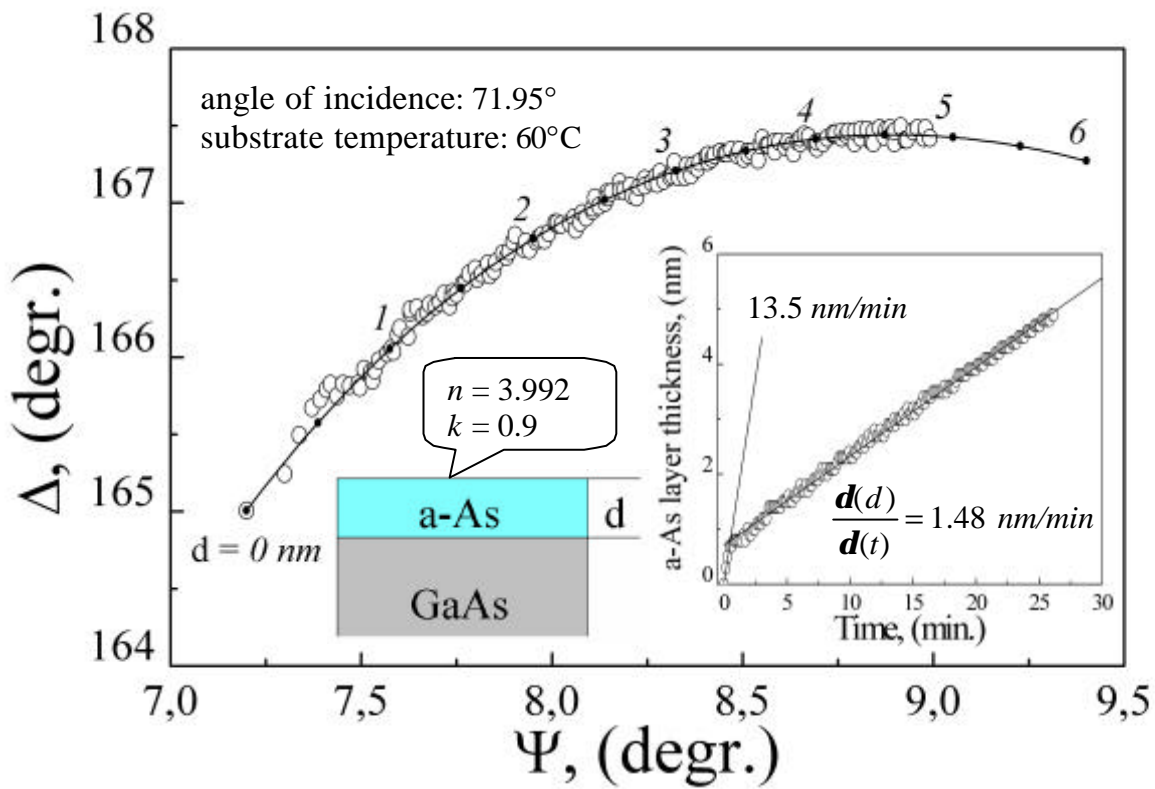

**Figure 2** Evolution of ellipsometric parameters $\psi$ and $\Delta$ during condensation of $As_4$ molecules on GaAs(001) surface. A theoretical curve fitting experimental points was calculated within one-layer model *GaAs–a-As–vacuum*. Insert shows time dependence of a-As layer thickness.

It has been shown in ref. [6] that GaAs samples prepared according to the procedure described above are covered by ~2 monolayers of chemisorbed amorphous arsenic. During vacuum heating this arsenic is desorbed with all surface contaminations on it being removed. It makes it possible to get atomically clean GaAs surface at temperatures as low as 400÷420°C [6-7]. After heating the samples to 540°C and subsequent cooling down to 60°C we have obtained atomically clean GaAs surface with Ga-stabilized reconstruction (4×6) as determined by RHEED. Then the shutter of As beam source was opened and condensation proceeded.

Fig.2 shows the evolution of ellipsometric parameters $\psi$ and $\Delta$ during growth of a-As layer from $As_4$ beam with equivalent pressure $BEP(As_4)=3\times10^{-4}$ torr at the substrate temperature 60°C. Experimental data are fitted by theoretical curve calculated within one-layer model *GaAs–a-As–vacuum*. The fit was accomplished by least squares method using a-As refractive ($n$) and absorption ($k$) indexes as two variable parameters for prescribed a-As film thickness in the range 0÷6nm. The best result was obtained for $n=3.992$ and $k=0.9$. Using these values the time dependence of a-As layer thickness was determined which is shown in the insert (Fig.2). As can be seen there is a kink on the kinetic curve at the film thickness of ~0.8nm. Since substrate temperature and $As_4$ beam intensity were constant during experiment the kink occurrence could be traced to different adsorption rates of incoming $As_4$ molecules. Up to $d=0.8$nm arsenic interacts with top monolayer of Ga atoms covering the substrate surface. AFM data (Fig.3) indicate that mean roughness of substrate is ~0.85nm and one could reasonably assume that formation of strong Ga–As bonds takes place until all unsatisfied Ga orbitals are saturated. Afterwards the sticking of $As_4$ molecules to the surface is governed by the formation of weaker As–As bonds.

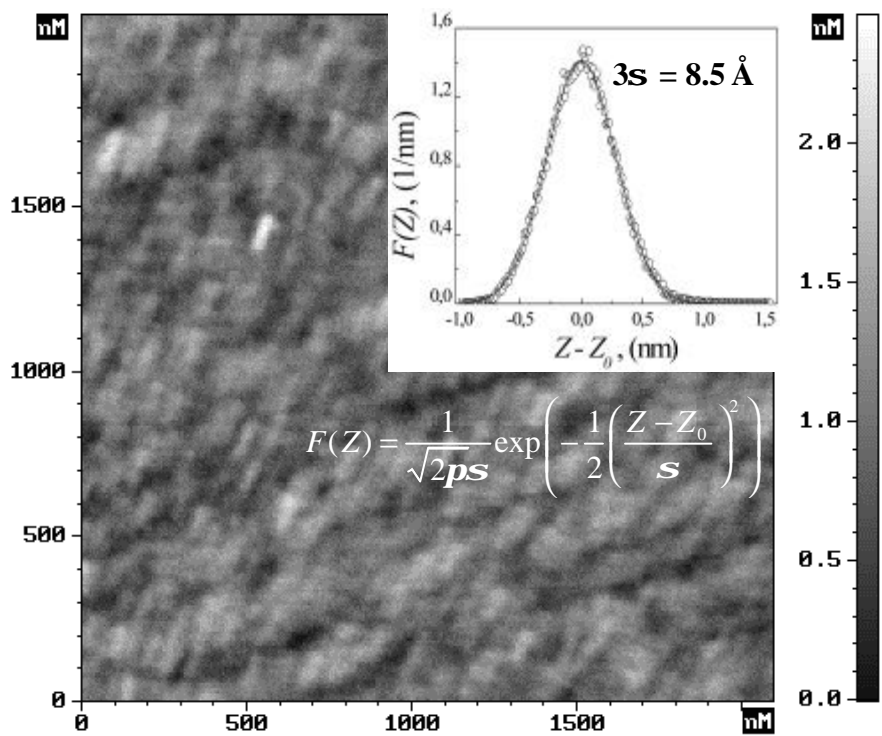

**Figure 3**. AFM picture of the GaAs(001) substrate surface used in condensation experiment. Insert shows height distribution over the scanned area $F(Z)$.

From our experimental data the sticking coefficient of $As_4$ molecules can be estimated. Tetrametric arsenic beam equivalent pressure $BEP(As_4)=3\times10^{-4}$ torr corresponds to the molecule flux density of

$$J(As_4) = \frac{BEP(As_4)}{\mathbf{h}(As_4) \cdot T_{guage}} \cdot \sqrt{\frac{8 \cdot T_{source}}{\mathbf{p} \cdot k \cdot m(As_4)}} \cong 3.1\times10^{16} \text{ molecules/cm}^2 \cdot \text{sec}, \qquad (1)$$

were $\mathbf{h}(As_4)=6{,}8\pm0{,}2$ is the sensitivity coefficient of ionization gauge for $As_4$ molecules [8], $T_{guage}=300K$ is temperature of the gauge, $T_{source}=680K$ is arsenic source temperature, $m(As_4)$ is the mass of the molecule, $k$ is Boltzmann constant. If all incoming molecules were sticking to the surface the growth rate would be

$$\frac{\mathbf{d}(d)}{\mathbf{d}(t)} = J(As_4) \cdot 4 \cdot V(As) \cong 1.88 \,\mu\text{m/min}, \qquad (2)$$

where $V(As) \sim 2.5\times10^{-23}\text{cm}^3$ is atomic volume of As in the films. In fact the growth rate is much lower and is about 1.48nm/min for $d>0.8$nm (see insert in Fig.2). It means that sticking coefficient of $As_4$ molecules to the surface of a-As film is only $\sim 8\times10^{-4}$ at 60°C. For Ga-rich substrate the growth rate is almost one order of magnitude higher and $As_4$ sticking coefficient is about $7\times10^{-3}$.

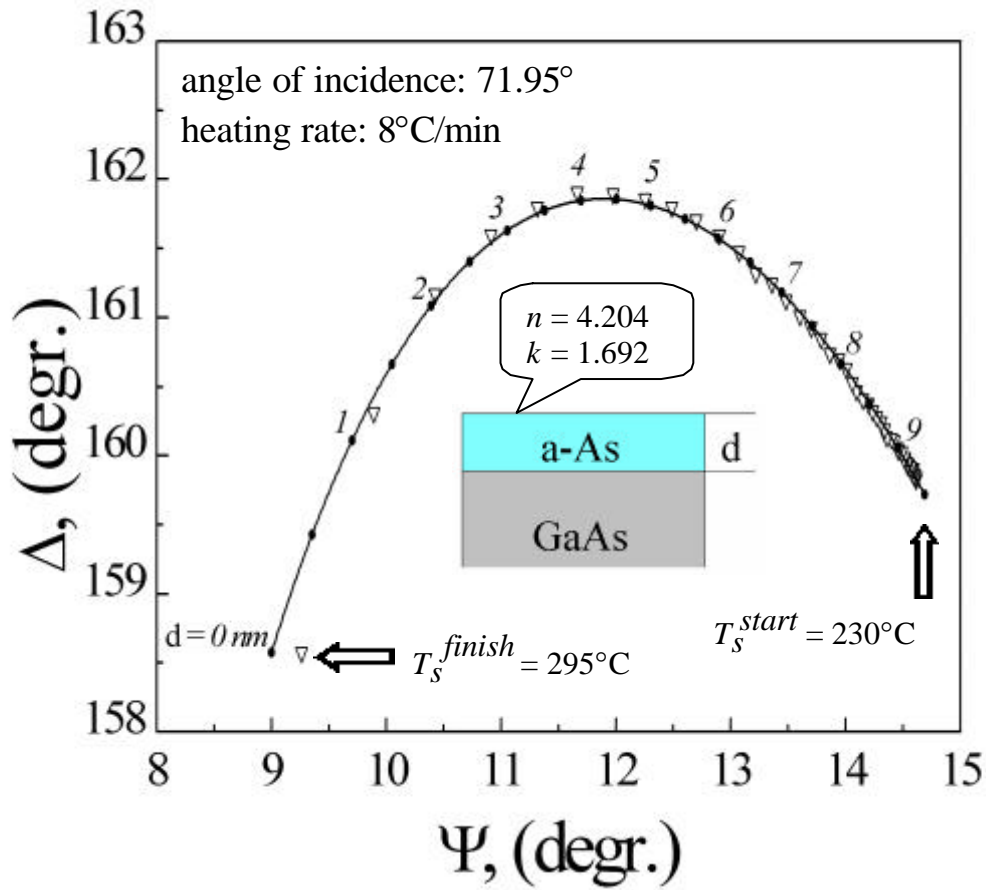

**Figure 4**. Evolution of ellipsometric parameters $\psi$ and $\Delta$ during sublimation of a-As film. A theoretical curve fitting experimental points was calculated within one-layer model *GaAs–a-As–vacuum*.

**SUBLIMATION**

For sublimation studies we have used undoped MBE-growth GaAs layers (~70nm thick) capped by a-As thin films directly after growth. Cap layer was formed by condensing the $As_2$ molecular beam from cracker cell on the as-grown sample held at sufficiently low temperature. The transfer of samples from the growth chamber to our experimental set-up as well as their mounting onto Mo holder was performed in ambient conditions. Fig.4 shows the evolution of ellipsometric parameters $\psi$ and $\Delta$ (open triangles) during sublimation of a-As film in the course of vacuum heating from 230°C up to 295°C with a rate of 8°C/min. Also shown in the figure are the results of theoretical fit performed in the framework of one-layer model for the system *GaAs–a-As–vacuum* (see insert). It can be seen that for *n*=4.204 and *k*=1.692 excellent agreement is obtained between experimental data and model calculations. This testifies to correctness of the model used. Several important conclusions can be drawn from the results obtained. Firstly, the sublimation of a-As film proceeds in planar mode, i.e. with decreasing film thickness during heating the sample surface remains smooth and additional roughness is not developed. Secondary, the rate of change in sample optical properties is strongly temperature dependent, which is evidenced by increasing distances between experimental points during heating. From measured results we have calculated the dependence of a-As layer thickness on the time (temperature), as shown in Fig.5. It should be also mentioned that optical properties of a-As layers grown from $As_4$ and $As_2$ molecules are very different. Absorption index for the film obtained from $As_2$ beam is a factor of two higher than that for As$_4$-grown layer. This difference can not explained by the temperature dependence of optical constants. It seems more natural to assume that using different building blocks we obtain structurally dissimilar a-As films.

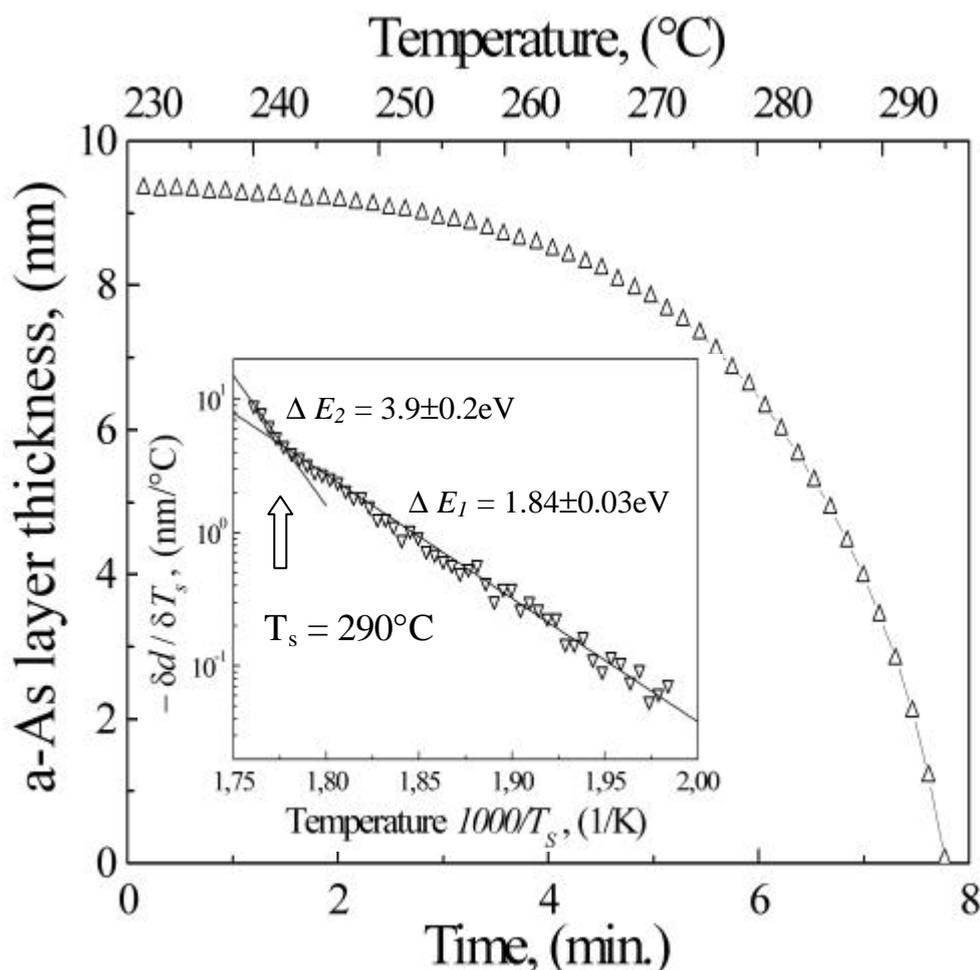

**Figure 5**. Time (temperature) dependence of a-As layer thickness during sublimation. Insert shows the sublimation rate versus inverse substrate temperature.

Since a-As layer thickness variation during heating can be written as

$$d(T_s) = D(T_s^{start}) - \frac{4 \cdot V(As)}{g} \int_{T_s^{start}}^{T_s} J(T_s') dT_s', \quad (3)$$

where $J(T_s)$ is the flux density of As$_4$ molecules from the surface at running temperature $T_s$, $D$ is the initial thickness of the layer at $T_s^{start}$ and $g$ is the heating rate. One obtains

$$-\frac{d(d)}{d(T_s)} = \frac{4 \cdot V(As)}{g} J(T_s) \quad (4)$$

According to Hertz–Knudsen–Langmuir equation the flux density of subliming molecules is given by

$$J(T_s) = \frac{a \cdot P(T_s)}{\sqrt{2p \cdot m(As_4) \cdot k \cdot T_s}}, \quad (5)$$

where $a$ is the evaporation coefficient and $P(T_s)$ is the equilibrium vapor pressure of As$_4$ molecules over solid arsenic at the given temperature. Assuming that

$$a \cdot P(T_s) \propto \exp\left(-\frac{\Delta E}{kT_s}\right), \quad (6)$$

where $\Delta E$ is the activation energy of sublimation and substituting (5) and (6) into (4) one obtains

$$-\frac{d(d)}{d(T_s)} \propto \frac{1}{\sqrt{T_s}} \exp\left(-\frac{\Delta E}{kT_s}\right). \quad (7)$$

Since in the temperature range 230÷295°C the factor $T_s^{-1/2}$ varies insignificantly expression (7) can be rewritten in the form

$$-\frac{d(d)}{d(T_s)} \propto \exp\left(-\frac{\Delta E}{kT_s}\right). \quad (8)$$

As a result, plotting the logarithm of sublimation rate versus inverse temperature one should obtain a straight line. Insert in Fig.5 shows the corresponding curve from which the activation energy for sublimation can be extracted. It is clearly seen that sublimation process of a-As film is a complicated phenomenon. In the temperature interval 230÷290°C the process is characterized by activation energy of $\Delta E_1$=1.84±0.03eV whereas the final part of sublimation (at 290÷295°C) requires a much higher activation energy $\Delta E_2$=3.9±0.2eV. We suggest that such a behaviour can be taken as an indication of some structural rearrangement in amorphous As at ~290°C. Possibly the As atoms in ultrathin (~3÷4nm) amorphous film relax to more stable configuration at this temperature and it takes more energy to disrupt the newly formed structure.

According to [9] the enthalpy of *solid–vapor* phase transition at 200÷300°C is $\Delta H^{sol-vap}$= 1.45eV. It means that for detachment of As$_4$ molecule from the surface of a-As layer additional energy barrier exist $\Delta E^a = \Delta E_1 - \Delta H^{sol-vap}$=0.39eV. This barrier will determine the dependence of evaporation coefficient α through eq. (6).

**ACKNOWLEDGMENTS**

partly supported by Ministry of Education of Russian Federation through the Task Program "Integration" (Contract № 0765/785).

**REFERENCES**
1. Resch U., Scholz W., Rossow U., Muller A.B., Richter W. and Forster A., Thermal desorption of amorphous arsenic caps from GaAs(001) monitored by reflection anisotropy spectroscopy, Appl. Surf. Sci., 1993, Vol.63(1-4), P.106-110.
2. Bernstein R.W., Borg A., Husby H., Fimland B.-O. and Grepstad J.K., Capping and decapping of MBE grown GaAs(001), $Al_{0.5}Ga_{0.5}As$(001) and AlAs investigated by ASP, PES, LEED and RHEED, Appl. Surf. Sci., 1992, Vol.56-58, P.74-80.
3. Gallagher M.C., Prince R.H. and Willis R.F., On the atomic structure and electronic properties of decapped GaAs(001) (2×4) surface, Surf. Sci., 1992, Vol.575(1-4), P.31-40.
4. Vitomirov I.M., Raisanen A.D., Finnefrock A.C., Viturro R.E., Brillson L.J., Kirchner P.D., Pettit G.D. and Woodal J.M., Temperature-dependent chemical and electronic structure of reconstructed GaAs(001) surface, J. Vac. Sci. Technol. B, 1992, Vol.10(4), P.1898-1903.
5. Vasquez R.P., Lewis B.F., Grunthaner F.J., Cleaning chemistry of GaAs(001) and InSb(001) substrates for molecular beam epitaxy, J. Vac. Sci. Technol. B, 1983, Vol.1(3), P.791-794.
6. Tereshchenko O.E., Chikichev S.I., Terekhov A.S., Composition and structure of HCl-isopropanol treated and vacuum annealed GaAs(001) surfaces, J. Vac. Sci. Technol. A, 1999, Vol.17(5) P.2655-2662.
7. Galitsin Yu.G., Mansurov V.G., Poshevnev V.I., Terekhov A.S., GaAs surface passivation by HCl alcohol solution, Poverhnost', 1989, № 10, P.140-142 (in Russian).
8. Preobrazhenskii V.V., Putyato M.A., Pchelyakov O.P., Semyagin B.R., Experimental determination of the incorporation factor of $As_4$ during molecular beam epitaxy of GaAs, J. Crystal Growth, 1999, Vol.201-202, P.170-173.
9. *The Characterization of High Temperature Vapors*, ed. Margrave J., John Wiley & Sons (New York, 1967).